\documentclass[
aps,
prl,
reprint,
superscriptaddress,
amsfonts,
amsmath,
amssymb,longbibliography
]{revtex4-2}

\usepackage{graphicx}
\usepackage{dcolumn}
\usepackage{bm}
\usepackage{hyperref}
\usepackage[mathlines]{lineno}
\usepackage{siunitx}
\usepackage{amssymb}
\usepackage{amsmath}
\usepackage{mathtools}
\usepackage[version=4]{mhchem}
\usepackage{multirow}
\usepackage{booktabs}
\usepackage{xcolor}
\usepackage{comment}
\usepackage{float}

\newcommand{\ltrap}{\textsc{Liontrap}}

\newcommand{\uw}{\textsc{uw}}
\newcommand{\katrin}{\textsc{katrin}}
\newcommand{\fsu}{\textsc{fsu}}
\newcommand{\alphatrap}{\textsc{alphatrap}}

\newcommand{\Cfive}{\ce{^{12}C^{5}+}}
\newcommand{\Csix}{\ce{^{12}C^{6}+}}
\newcommand{\Catom}{\ce{^{12}C}}
\newcommand{\Heatom}{\ce{^{4}He}}
\newcommand{\Hethreeatom}{\ce{^{3}He}}
\newcommand{\Hethreeone}{\ce{^{3}He^+}}

\newcommand{\Hetwo}{\ce{^{4}He^{2}+}}
\newcommand{\Heone}{\ce{^{4}He^+}}
\newcommand{\hd}{\ce{HD^+}}
\newcommand{\htwo}{\ce{H_2^{+}}}

\DeclareSIUnit\atomicmassunit{u}
\DeclareSIUnit\bar{bar}

\begin{document}

\title{Penning-Trap Mass Measurement of Helium-4}

\author{S.~Sasidharan}
\affiliation{Max-Planck-Institut für Kernphysik, Saupfercheckweg 1, 69117 Heidelberg, Germany}
\affiliation{GSI Helmholtzzentrum für Schwerionenforschung GmbH, Planckstraße 1, 64291 Darmstadt, Germany}
\affiliation{Heidelberg University, Grabengasse 1, 69117 Heidelberg, Germany}
\author{O.~Bezrodnova}
\affiliation{Max-Planck-Institut für Kernphysik, Saupfercheckweg 1, 69117 Heidelberg, Germany}
\author{S.~Rau}
\affiliation{Max-Planck-Institut für Kernphysik, Saupfercheckweg 1, 69117 Heidelberg, Germany}
\author{W.~Quint}
\affiliation{GSI Helmholtzzentrum für Schwerionenforschung GmbH, Planckstraße 1, 64291 Darmstadt, Germany}
\author{S.~Sturm}
\affiliation{Max-Planck-Institut für Kernphysik, Saupfercheckweg 1, 69117 Heidelberg, Germany}
\author{K.~Blaum}
\affiliation{Max-Planck-Institut für Kernphysik, Saupfercheckweg 1, 69117 Heidelberg, Germany}


\begin{abstract}
 Light-Ion Trap ($\ltrap$), a high-precision Penning-trap mass spectrometer, was used to determine the atomic mass of $\Heatom$. Here, we report a 12 parts-per-trillion measurement of the mass of a $\Hetwo$ ion, $m$($\Hetwo$) = \SI[parse-numbers=false ]{$4.001\:506\:179\:651$(48)}{\atomicmassunit}. From this, the atomic mass of the neutral atom can be determined without loss of precision: $m$($\Heatom$) = \SI[parse-numbers=false]{$4.002\:603\:254\:653$(48)}{\atomicmassunit}. This result is slightly more precise than the current CODATA18 literature value but deviates by 6.6 standard deviations.\par
  \textit{This is a post-peer-review, pre-copy edit version of an article published in PRL. The final version is available online at}~\url{https://doi.org/10.1103/PhysRevLett.131.093201}.
 
\end{abstract}

\maketitle
Precision experiments at low energies have taken a pivotal role in the validation of the standard model of physics and potentially its boundaries. Experimental observables of simple atomic systems with a single or few electrons, such as atomic and molecular hydrogen, helium, and their ions, enable extremely precise tests. The properties of the fundamental particles, such as the mass of the electron~\cite{emass,HD_emass} and the mass and charge radius of the light nuclei~\cite{ed_atomicmasses_general,liontrap,charge_radius_He}, are thus of importance for fundamental physics. They interconnect a variety of individual observables from transition frequencies in hydrogen, ruled by the Rydberg constant~\cite{rydberg}, to the magnetic moment of the electron, which depends on the fine-structure constant~\cite{Electron_Magnetic_Moment}. Moreover, the mass difference between tritium and helium-3 can serve as a cross-check for the systematics in the search for a nonzero electron antineutrino mass with the $\katrin$ experiment, which studies the endpoint energy region of the $\beta$-decay spectrum of tritium~\cite{katrin}.\par
Today, the ratio of atomic masses can be measured to parts-per-trillion (ppt) relative precision using specialized Penning traps~\cite{FSUionbalance,Pentatrap_most_precise,deuteron}. While such measurements are largely independent of theory and generally reliable, the mass ratios of light ions have seen inconsistencies in the past~\cite{liontrap,ed_atomicmasses_general}. The atomic masses of the proton and deuteron measured in our group~\cite{proton,liontrap,deuteron} have discrepancies with the results from the group of Van Dyck at the University of Washington ($\uw$)~\cite{UW_protonmass,vandyck3HeandD}, which served as previous literature values.  However, our results are in agreement with the precise ratios of $\htwo$ and deuteron measured at Florida State University ($\fsu$)~\cite{FSUdtop, FSUionbalance}, and the mass ratios extracted from the high-precision laser spectroscopy of the molecular $\hd$ ion ~\cite{sayanHDplus4,KortunovHD}. Hence, it appears that the disparities are likely caused by the $\uw$ results.\par
At present, the $\Heatom$ mass measured with a precision of 16 ppt at $\uw$ solely yields the accepted literature value by Committee on Data for Science and Technology (CODATA) and Atomic Mass Evaluation (AME)~\cite{vandyck2006,codata2018,ame2020}. This is not ideal, especially in light of an increasing interest in helium for fundamental physics: from the developments in ultra-precision laser spectroscopy on $\Heone$ ions~\cite{He_electronic_spectroscopy} or antiprotonic helium~\cite{antiprotonichelium} to an improved determination of the electron’s atomic mass~\cite{emassHe}, the anticipated advancements necessitate a reliable $\Heatom$ mass value. With state-of-the-art Penning trap techniques, a significant improvement of the electron’s atomic mass via a measurement of the bound-electron $\it{g}$-factor in hydrogen-like ions should be possible. Such $\it{g}$-factor determination to a relevant precision is achievable and planned by setups like $\alphatrap$~\cite{alphatrap_tim} and the experiment that recently measured the $\Hethreeone$ magnetic moment~\cite{3Hegfactor}. This paves the way for potential progress in determining the mass of the electron in the near future. The current CODATA18 value comes from a measurement with $\Cfive$~\cite{emass,codata2018}. While a measurement with $\Catom$ directly yields a result in atomic mass units (u), the relatively high charge state increases the impact of higher-order contributions of quantum electrodynamics on the theoretical value of the $\it{g}$-factor. Consequently, a measurement with $\Heone$, where these terms are negligible, could potentially increase the precision and reliability of the electron’s atomic mass~\cite{emassHe}. To support such an electron mass determination, a consistent and precise $\Heatom$ mass value is required. Given the observed discrepancies with the $\uw$ results and the inconsistencies found in the mass measurements of $\Heatom$ from different groups~\cite{vandyck1995,Mainz2001,smiletrap2001,vandyck2004,vandyck2006,smiletrap2006}, it seems imperative to remeasure the $\Heatom$ mass.\par
The $\ltrap$ experiment is a high-precision Penning-trap mass spectrometer~\cite{liontrap}, where an ion is confined radially by a homogeneous magnetic field and axially by an electrostatic quadrupole potential ~\cite{gabrielse_geoniumtheory,dehmelt}. The ion exhibits three eigenmotions with corresponding frequencies: the axial motion along the magnetic field with frequency $\nu_z$, the modified cyclotron motion, and the magnetron motion in the perpendicular radial plane with frequencies $\nu_+$ and $\nu_-$, respectively. The free cyclotron frequency of the ion, 
\begin{equation} \label{1}
\begin{aligned}
\nu_c=\frac{1}{2\pi}\frac{q}{m} B,
\end{aligned}
\end{equation}
\begin{figure*}[ht]
\includegraphics[width=16.2cm, height=6.9cm]{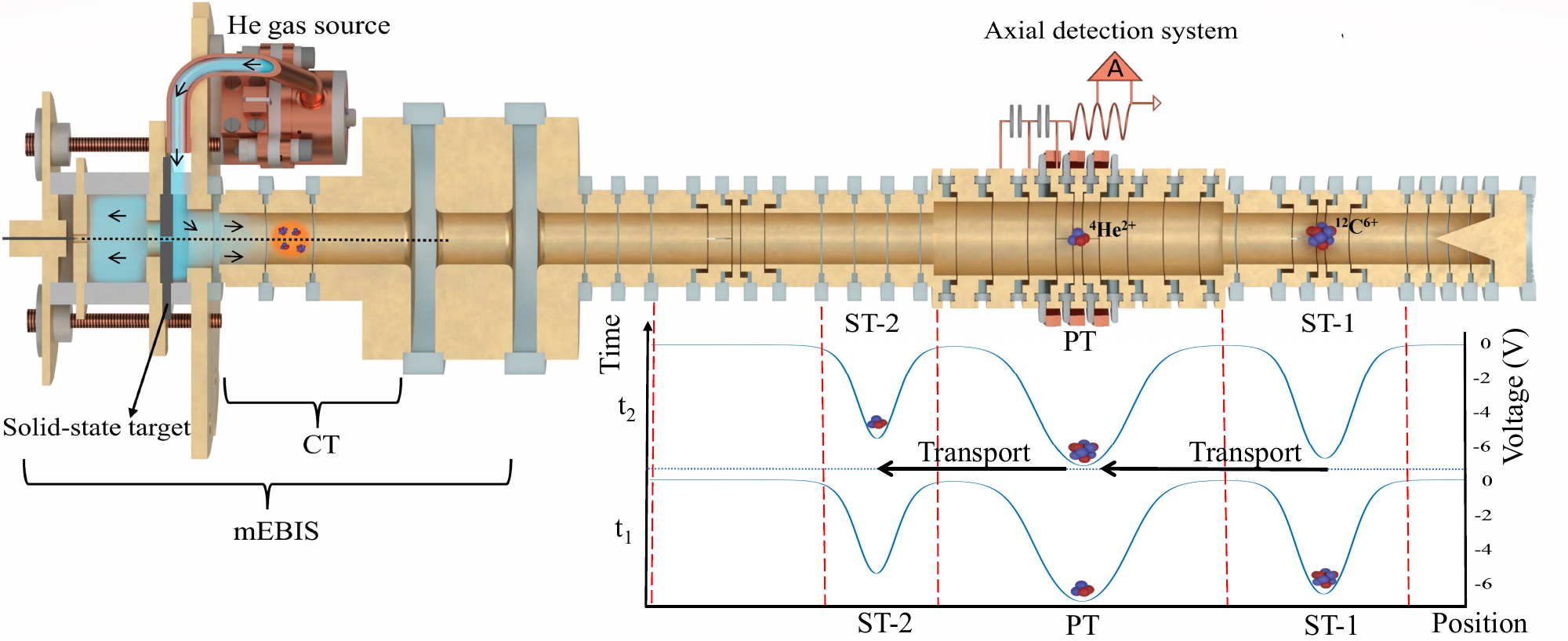}
\caption{\label{Trap stack} Cross-sectional sketch of the trap setup. The mEBIS produces ions in situ. $\Heatom$ atoms (blue) are released upon heating the source and guided into the electron beam (dotted black line) from the FEP, which enables the production of $\Heatom$ ions. The electron beam also ablates and ionizes atoms from the target's surface. The trapped ion cloud (orange-shaded spot) is adiabatically transported from the CT to the PT to prepare single ions. The axial detection system in the PT is also depicted. The lower right side of the image illustrates how the ions are stored and shuttled between the PT and one of the STs for the $\nu_c$ measurements. In this example, $\Hetwo$ and $\Csix$ are measured at time t$_1$ and t$_2$, respectively.}
\end{figure*}
where $q$ is the charge and $m$ is the mass of the ion in a magnetic field $B$, can be calculated from the eigenfrequencies using the invariance theorem \cite{gabrielse}:
\begin{equation} \label{2}
\begin{aligned}
    \nu_c = \sqrt{{\nu_+}^2+{\nu_-}^2+{\nu_z}^2}.
\end{aligned}
\end{equation}

The mass measurement technique is based on comparing cyclotron frequencies of the ion of interest and a reference ion in the same magnetic field, allowing the cancellation of the magnetic field in Eq.~(\ref{1}). In this work, we choose a carbon $\Csix$ ion as a reference for a $\Hetwo$ ion, which allows measurement in atomic mass units (u). 
The mass of an atom is related to the mass of its ion via the ionization energies~\cite{NIST} and the mass of the missing electrons without relevant loss of precision in the low $Z$ regime. The mass of $\Hetwo$ can be expressed as 
\begin{equation} \label{3}
\begin{aligned}
m(\Hetwo)=\frac{2}{6}\frac{\nu_c(\Csix)}{\nu_c(\Hetwo)} m(\Csix).
\end{aligned}
\end{equation}\par
A hermetically sealed trap chamber and the cryoelectronics are placed in the homogeneous field of a \SI{3.8}{\tesla} superconducting magnet and cooled to \SI{4.2}{\kelvin}. The vacuum inside the trap chamber is better than \SI[print-unity-mantissa = false]{e-17}{\milli\bar}, which allows several months of ion storage time. The trap chamber contains a stack composed of the Precision Trap (PT), two Storage Traps (ST-1 and ST-2), several transport electrodes, and a miniature electron beam ion source (mEBIS) that includes a Creation Trap (CT) and a Field-Emission-Point (FEP) electron source (see Fig.~\ref{Trap stack}). The measurements are performed in the PT, which is a doubly compensated, seven-electrode cylindrical Penning trap~\cite{liontrap}.\par
The mEBIS is used to produce ions \textit{in situ}. The carbon atoms are ablated from a solid-state target by the electron beam of the mEBIS~\cite{deuteron}. However, due to weak bonding capabilities, this technique does not work for noble gas atoms. Hence, a source was developed where gas atoms are adsorbed at \SI{4}{\kelvin} on a large surface area, which upon heating, releases the gas required for ionization and trapping (see Fig~\ref{Trap stack}). The source originally planned for $\Hethreeatom$ production was repurposed to produce $\Heatom$ atoms due to a technical issue. A future publication is planned to present a description and characterization of the source.\par
\begin{figure}[t]
\centering
\includegraphics[width=8.6cm, height=5cm]{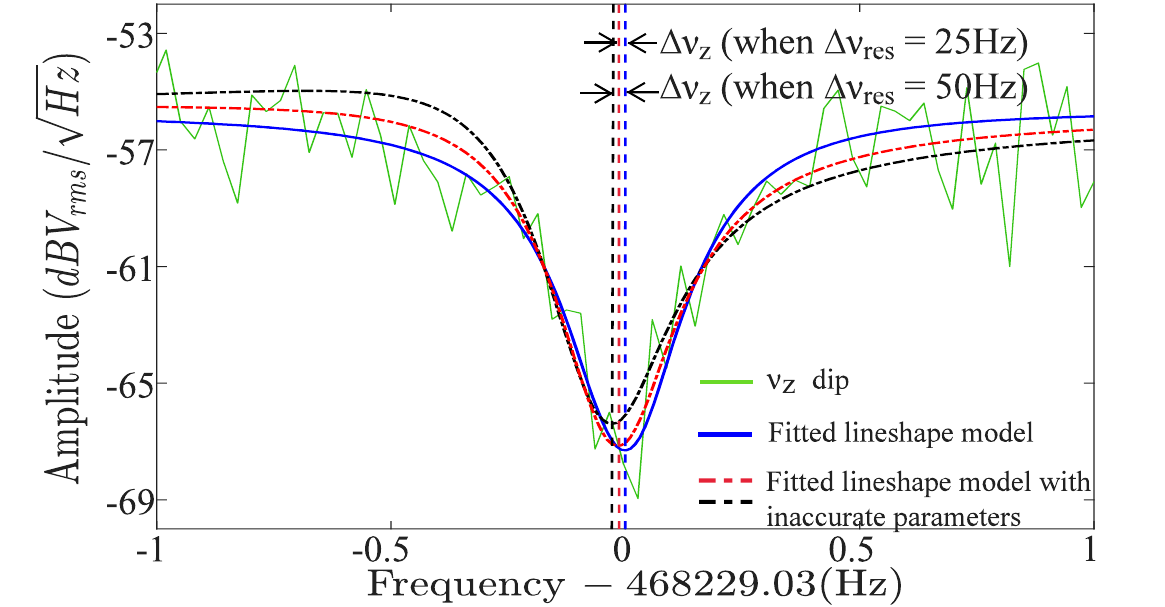}
\caption{\label{Dip} An axial dip spectrum of $\Hetwo$. The axial frequency of the ion is extracted from the $\sim$ \SI{500}{\milli\hertz} wide ion signal (green) fit with a lineshape model (blue). To this end, the resonator's parameters must be precisely known. The black and red lines indicate the lineshape that results from artificially shifting the resonator frequency by \SI{50}{Hz} and \SI{25}{Hz}, respectively. Due to the resulting asymmetry, these fits result in systematically shifted axial frequency values (marked with vertical dotted lines).}
\end{figure}

  
After production, the $\Csix$ and $\Hetwo$ ions are stored simultaneously within the trap stack in separate potential minima. The cyclotron frequencies are measured alternately in the PT~\cite{liontrap}, as shown in the inset of Fig.~\ref{Trap stack}. For both $\nu_c$ measurements, the same trapping potential is applied to ensure the ions are at the identical position in the trap. The ion's axial motion is detected via the image currents ($\sim$\SI[print-unity-mantissa = false]{e-15}{\ampere}) induced on the trap electrodes by its harmonic axial oscillations. A tank circuit (resonator) connected to these electrodes converts the image current into a measurable voltage signal ($\sim$ \SI[print-unity-mantissa = false]{e-9}{\volt}) and cools the ion's axial motion to \SI{4.2}{\kelvin}. The resonator's frequency is tuned with a varactor diode to get it in resonance with either one of the two ions without altering the trapping potential. In equilibrium, the thermal noise of the resonator gets electrically shorted, producing a minimum in the noise spectrum at the ion's axial frequency, known as ``dip'', Fig.~\ref{Dip}. While this technique allows measurement at low axial energy, it has the drawback that the interaction between the ion and the resonator causes frequency-pulling~\cite{vasant_frequency_pulling}. Thus, a lineshape model must be fitted to the dip to extract the unperturbed ion frequency~\cite{liontrap}. At a trapping voltage of typically \SI{-7.8}{\volt}, the frequencies of the ions with charge-to-mass ratio (\(\frac{q}{m}\)) = \(\frac{1}{2}\) \(\frac{e}{u}\) are of order $\nu_z\approx$ \SI{468}{\kilo\hertz}, $\nu_+\approx$ \SI{29}{\mega\hertz} and $\nu_-\approx$ \SI{4}{\kilo\hertz}. The hierarchy between the frequencies entering the invariance theorem is evident as $\nu_+ \gg\nu_z \gg\nu_-$, implying the modified cyclotron frequency needs to be determined with the highest precision.\par
In $\ltrap$, we can use the phase-sensitive Ramsey-like measurement technique PnA (``Pulse and Amplify'')~\cite{PnA} and the ``double-dip'' method to extract the modified cyclotron frequency~\cite{liontrap}.
To avoid lineshape-related problems, we use the PnA technique to extract the modified cyclotron frequency with high precision. In this technique, an excitation pulse with frequency $\nu_+$ is initially applied to excite and imprint a phase on the cyclotron mode.
After a certain evolution time T$_{evol}$, the resulting phase of the modified cyclotron motion is transferred to the axial motion by a parametric amplification pulse at the sideband $\nu_+ + \nu_z$. The axial phase is read from the Fourier transform of the image current signal to determine the modified cyclotron frequency. The magnetron frequency is measured only occasionally during the measurement campaign using the double-dip technique~\cite{liontrap}.\par
\begin{figure}[t] 
\centering
\includegraphics[width=8.6cm, height=3.7cm]{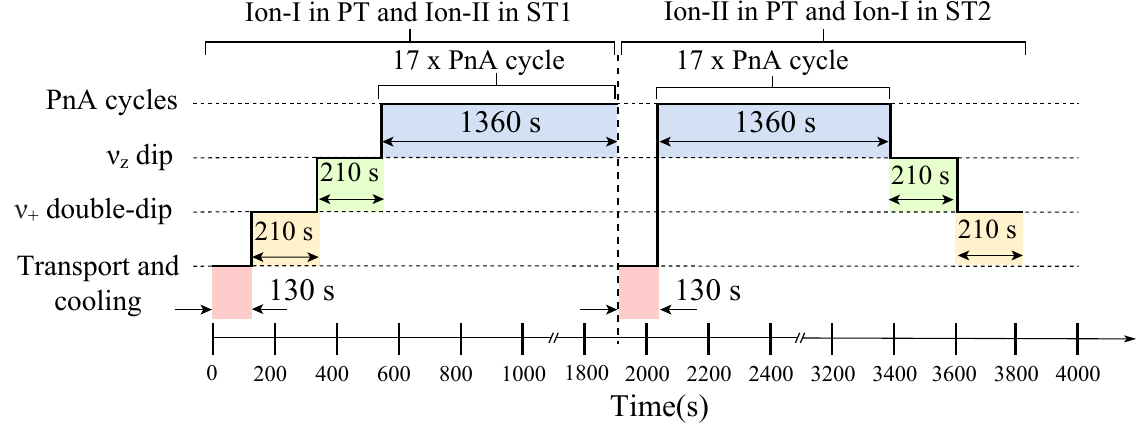}
\caption{\label{Measurement cycle} Timeline diagram of a measurement cycle. At the beginning of a cycle, ion-I is selected randomly and transported to PT, where all the modes are cooled, and the $\nu_z$ (dip) and $\nu_+$ (double-dip) are measured, followed by a $\nu_+$~(PnA) measurement. Ion-I is then moved away, ion-II is transported to the PT, and the frequencies are measured in the inverse sequence. The time periods mentioned are only approximate. Every measurement cycle consists of 17 PnA cycles, each performed with a different T$_{evol}$: 6 $\times$ \SI{0.1}{\second}, \SI{0.5}{\second}, \SI{1}{\second}, \SI{2}{\second}, \SI{5}{\second}, \SI{10}{\second}, 6 $\times$ \SI{20}{\second}. The long T$_{evol}$ enable the determination of $\nu_+$ with the highest precision.}
\end{figure}
\begin{figure}[t]
\centering
\includegraphics[width=8.6cm, height=4.8cm]{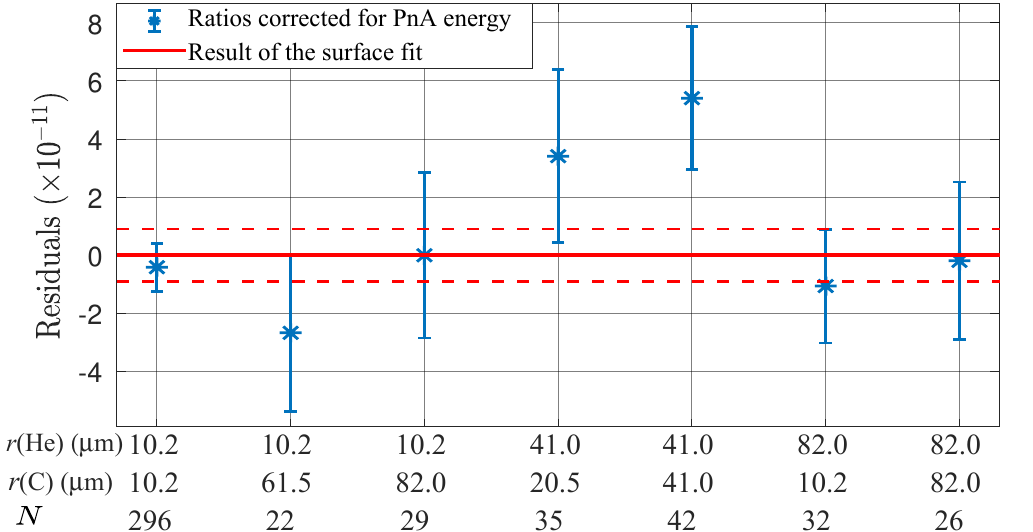}
\caption{\label{surface fit} Residuals of the 3-parameter surface fit. Individual points in the plot correspond to different PnA settings. The $x$-axis marks the respective cyclotron radii of $\Hetwo$ $r$(He) and $\Csix$ $r$(C) and the number of cyclotron frequency ratios used to obtain each value ($N$). Each point is the mean of the corresponding $N$ measurements, and the error bar is the standard error of the mean. The $y$-axis shows the residuals of the surface fit for different excitation settings. The region between the dotted lines indicates the 1$\sigma$ confidence interval of the frequency ratio extracted from the fit ($R^{CF}_\mathrm{stat}$). }
\end{figure}

The techniques discussed above are put together to perform a mass measurement cycle summarised in Fig.~\ref{Measurement cycle} and detailed in~\cite{liontrap}. Each cycle produces a cyclotron frequency ratio:
\begin{equation} \label{4}
\begin{aligned}
R^{CF}= \frac{\nu_c(\Csix)}{\nu_c(\Hetwo)}.
\end{aligned}
\end{equation}
There were 32 runs with a total of 482 $R^{CF}$ measurements using two ion pairs stored in different spatial order to avoid any systematic effects originating from the order of ions or unnoticed contaminant ions. During different runs, the strength $S{_{t,\hat{U},i}^+}$ of the modified cyclotron frequency excitation and consequently the ion's cyclotron amplitude was varied. Here, $\hat{U}$ and $t$ are the amplitude and duration of the excitation pulse, respectively, and the different excitation settings are denoted as $\it{i}$. These runs are sorted based on $\it{i}$. The excitation strengths are related to the excited modified cyclotron radii as $r{_{i,exc}^+}=\kappa S{_{t,\hat{U},i}^+}$, where $\kappa$ is a calibration constant. The $r{_{exc}^+}$ of both ions were varied over a range of \SI{10}{\micro\meter} to \SI{80}{\micro\meter} between the runs resulting in different relativistic shifts of the cyclotron frequency of the ions. This shift is treated using a three-parameter surface fit that relates the individually shifted measurements $R{^{CF}_i}$ to $R{^\mathrm{CF}_\mathrm{stat}}$, which is the frequency ratio extrapolated to zero cyclotron excitation energy (Fig.~\ref{surface fit}):
\begin{equation} \label{5}
\scalebox{0.96}{$R{^{CF}_i} = R{^{CF}_\mathrm{stat}} + a{[S{_{t,\hat{U},i}^+}(\Hetwo)]}^2 + b{[S{_{t,\hat{U},i}^+}(\Csix)]}^2$}.
\end{equation}
The result of the surface fit is
\begin{equation} \label{6}
\begin{aligned}
R{^{CF}_\mathrm{stat}} = 1.000\;650\;921\;128\;8(90). 
\end{aligned}
\end{equation}
The $R{^{CF}_\mathrm{stat}}$ still needs to be corrected for further systematic shifts. These frequency shifts are summarized in Table ~\ref{Table of systematics}.\par
While the excitation radius is treated using the surface fit, the finite thermal radius of the ion leads to a residual relativistic shift in the cyclotron frequency. Before every PnA cycle, the ion’s eigenmodes are thermalized with the resonator. To minimize the shift, electronic feedback cooling is implemented to lower the axial temperature of the ion to $T_\mathrm{z,FB} =\:$\SI{1.7(3)}{\kelvin} ~\cite{sventhesis,florianthesis}. \par
The image charge shift caused due to the charges induced on the trap electrodes by the ion significantly contributes to the systematic error budget. The shift in cyclotron frequency depends on the trap design and is determined from numerical simulations and experimentally tested to a relative precision of 5$\%$~\cite{imagecharge}.\par
The PT is optimized to minimize the electrostatic anharmonicities described by even $C_i$-coefficients ($\it{i}\geq$ 4)~\cite{Ketter}. The trap was optimized to have $C_4=$ \num{0(1)e-5} and $C_6=$ \num{-4(15)e-5}, which lead to shifts less than 0.1 ppt in the cyclotron frequency ratio.\par
The residual magnetostatic inhomogeneity is characterized by the series expansion of the magnetic field along the trap axis ($B= B_0 + B_1z + B_2z^2$ +..). 
At $\ltrap$, a superconducting shim coil has been implemented around the trap chamber to minimize the quadratic component $B_2$ in situ as it is responsible for the first-order energy-dependent frequency shifts ~\cite{deuteron}. The residual $B_2$ = (\num{-0.4} $\pm$ 2.0) \si{\milli\tesla\per\square\meter} is small enough to make the frequency shift minor for the small motional amplitudes used in this work.\par
The lineshape model used to extract the unperturbed axial frequency from the dip signal has the resonance frequency $\nu_\mathrm{res}$ and the quality factor $Q$ of the resonator as input parameters, which are determined with a fit to the thermal noise of the resonator. If $\nu_\mathrm{res}$ or $Q$ are determined incorrectly, the extracted axial frequency is shifted, and this enters $\nu_{c}$ through the invariance theorem. Fig.~\ref{Dip} shows two exaggerated cases of incorrect $\nu_\mathrm{res}$. We assume $\nu_\mathrm{res}$ to be known to $\pm$ \SI{3}{\hertz} during the mass measurement, limited by the fit quality and fluctuations between runs, leading to a relative uncertainty of 7~ppt in the cyclotron frequency ratio.\par 
For each measurement cycle the magnetron frequency is calculated from the measured axial and modified cyclotron frequencies~\cite{gabrielse_geoniumtheory}, as it is only measured intermittently during the campaign. Although this calculated value is subject to systematic shifts from trap imperfections such as tilt and ellipticity~\cite{gabrielse}, it deviates by $\leq$ \SI{600}{\milli\hertz} from the measured value. Since this shift is almost independent of $q/m$, the effect on the cyclotron frequency ratio is $\sim10^{-15}$ and is thus not considered as a systematic error.\par 
\begin{table}[t]
\caption{\label{Table of systematics} Summary of the systematic shifts and their uncertainties for the cyclotron frequency ratio. $R{^{CF}_\mathrm{stat}}$ is the statistical result and $R{^{CF}_{corr}}$ is the ratio corrected for systematic shifts. All values are in parts-per-trillion and relative. The excitation amplitudes are extrapolated to zero.}
\centering
\begingroup
\begin{tabular}{l c c} 
\toprule
\multirow{1}{*}{Effect}& 
\multicolumn{1}{c}{\parbox{2.5cm}{Rel. shift in $R_{CF}$}}&
\multirow{1}{*}{\parbox{1.8cm}{Uncertainty}}\\
& $\left(\displaystyle\frac{R{^{CF}_\mathrm{stat}}-R{^{CF}_\mathrm{corr}}}{R{^{CF}_\mathrm{stat}}}\right)$ & \\ 
\midrule
Image charge & 65.76 & 3.29 \\ 
Relativistic effect & -1.77 & 0.31\\
Magnetic inhomogeneity & -0.03 & 0.19\\
Electrostatic anharmonicity & 0 & 0.13\\
Dip lineshape & 0 & 7.11\\ [1ex] 
\midrule
Total & 63.96 & 7.84 \\
\bottomrule
\end{tabular}
\endgroup
\end{table}
The ratio of cyclotron frequencies, considering both systematic and statistical uncertainties, is:
 \begin{equation} \label{7}
 \scalebox{0.99}{$R^{CF}_{Final} =1.000\;650\;921\;192\;9(90)_\mathrm{stat}(78)_\mathrm{sys}(119)_\mathrm{tot}$}.
\end{equation} 
From $R^{CF}_{Final}$ we can derive the mass of $\Hetwo$ via Eq.~(\ref{3}): 
\begin{equation} \label{8}
\scalebox{0.945}{$m(\Hetwo)= 4.001\;506\;179\;651(36)_\mathrm{stat}(31)_\mathrm{sys}(48)_\mathrm{tot}~\si{u}$}.
\end{equation} 
The numbers in brackets denote statistical, systematic, and total uncertainty, respectively. Considering the mass of the missing electrons~\cite{codata2018} and their binding energies~\cite{NIST}, the mass of the $\Heatom$ atom is calculated: 
\begin{equation} \label{9}
\scalebox{0.96}{$m(\Heatom)= 4.002\;603\;254\;653(36)_\mathrm{stat}(31)_\mathrm{sys}(48)_\mathrm{tot}~\si{u}$}. 
\end{equation} \par
\begin{figure}[H]
\centering
\includegraphics[width=8.6cm, height=11.2cm]{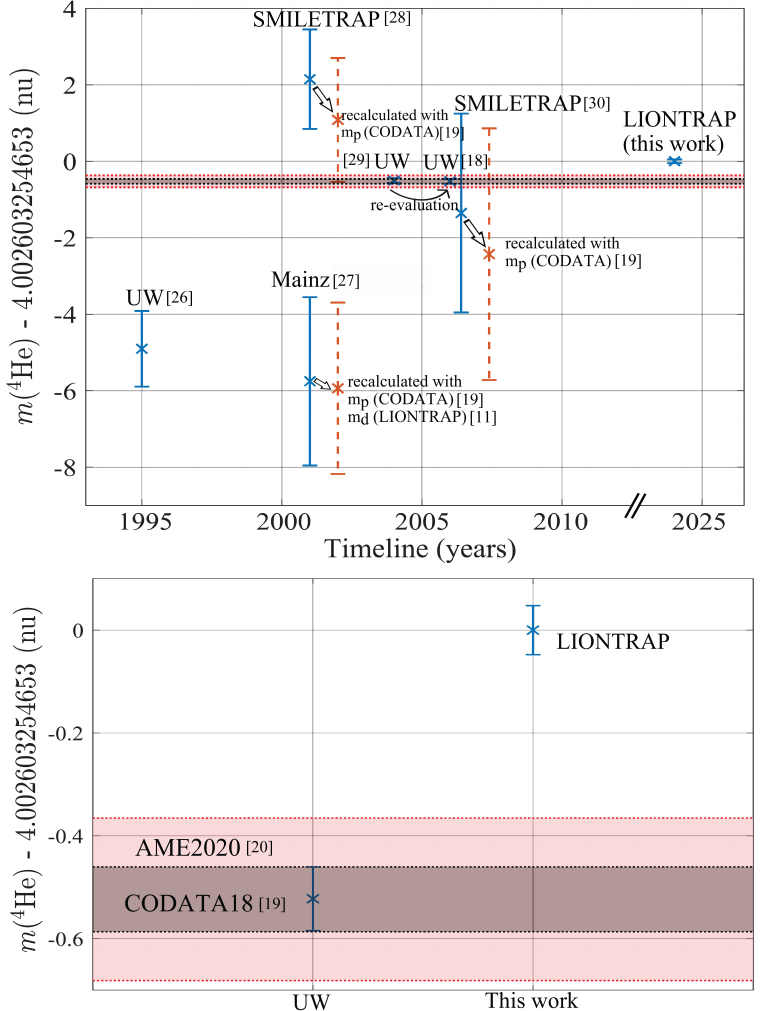}
\caption{\label{history plot} High-precision $\Heatom$ mass values from different experiments. On the top, $\Heatom$ mass measurements over the years~\protect{\cite{vandyck1995,vandyck2004,vandyck2006,Mainz2001,smiletrap2001,smiletrap2006}} are plotted as blue points. The red points indicate the mass of $\Heatom$ reevaluated based on the recent proton~\protect\cite{codata2018} or deuteron~\protect\cite{deuteron} mass values (where these were used as reference ions). The red shaded area indicates the AME2020 value, and the grey shaded region indicates the CODATA18 value. On the bottom, an enlarged into the result of this work is shown. The CODATA18 and AME2020 values coincide with the latest $\uw$ result (see text).}
\end{figure}

Compared to the CODATA / AME value, the $\Heatom$ mass reported in this work deviates notably~\cite{codata2018,ame2020}.
As previously stated, the CODATA18 value is given by the $\uw$ mass measurement of $\Heatom$ ion versus a $\Catom$ ion~\cite{vandyck2006}. The difference observed between this literature value and our $\Heatom$ mass value is about 6.6 combined standard deviations (Fig.~\ref{history plot}).\par
The literature value reported by the AME is identical to the one in CODATA18 but with an increased uncertainty~\cite{ame2020}. In AME2020~\cite{2ame2020}, the uncertainty for the mass of $\Heatom$ was increased from \SI{63}{\pico\atomicmassunit}~\cite{AME2016} to \SI{158}{\pico\atomicmassunit} due to the agreement between the results from the $\fsu$ trap and $\ltrap$ and the inconsistencies with $\uw$ results. Our result still deviates by 3.2 combined standard deviations from the AME2020 value. This discrepancy demands a re-measurement of the $\Hethreeatom$/$\Catom$ ~\cite{vandyck3HeandD} mass ratio, as planned in our experiment, to further investigate the inconsistencies in the light ion mass regime~\cite{deuteron}. Having noticed such deviations also for the masses of the proton and deuteron, several independent and consistent measurements are necessary to regain confidence in the masses of light ions. \par
\begin{acknowledgments}
This work was supported by the Max Planck Society. The project received funding from the R$\&$D cooperation agreement between GSI/FAIR and Heidelberg University, the International Max Planck Research School for Precision Tests of Fundamental Symmetries (IMPRS-PTFS) and from the Max Planck–RIKEN–PTB Center for Time, Constants and Fundamental Symmetries. This study comprises parts of the Ph.D. thesis work of S.~Sasidharan.
\end{acknowledgments}
\providecommand{\noopsort}[1]{}\providecommand{\singleletter}[1]{#1}%

\end{document}